# Social media cluster dynamics create resilient global hate highways

N.F. Johnson[1], R. Leahy[1], N. Johnson Restrepo[1], N. Velasquez[2], M. Zheng[3], P. Manrique[3]
[1]Physics Department, George Washington University, Washington D.C. 20052
[2]Elliot School of International Affairs, George Washington University, Washington D.C. 20052
[3]Physics Department, University of Miami, Coral Gables, Florida 33124

**Online social media allows individuals to cluster around common interests -- including hate. We show that tight-knit social clusters interlink to form resilient 'global hate highways' that bridge independent social network platforms, countries, languages and ideologies, and can quickly self-repair and rewire. We provide a mathematical theory that reveals a hidden resilience in the global axis of hate; explains a likely ineffectiveness of current control methods; and offers improvements. Our results reveal new science for networks-of-networks driven by bipartite dynamics, and should apply more broadly to illicit networks.**

Hate against a particular community has been expressed in one form or another for centuries [1,2]. Recent hate-driven violent attacks in the U.S. have targeted the black community (Jeffersontown 10/24/2018), politicians (nationally 10/25/2018), the Jewish community (Pittsburgh 10/26/2018), women (Tallahassee 11/2/2018) and even perhaps communities of students (California 11/8/2018). A significant recent development is that 'anti-X' narratives can now be developed and expressed through online social media [3,4]. Indeed, online hate has been frequently tied to extremism and violent real-world attacks including mass shootings [5,6,7]. Though strict definitions of hate and extremism may be hard to establish, many groups that have been labelled as hate groups are also labelled as extremist and vice versa [7,8,9]. There are many ongoing legal and academic debates as to exactly what online material represents hate, extremism or a national security threat -- yet in practice, social media companies such as Facebook are expected to aggressively eliminate such material from their platform in real time, while simultaneously balancing the right of free speech.

In addition to the difficult ongoing debates concerning content, the technical task of tracking online hate is an immense challenge. Each day, approximately 2.5 quintillion bytes of data are created by the 3+ billion worldwide Internet users [10] with 4 million hours of content uploaded to YouTube, 67,305,600 new Instagram posts, and 4.3 billion new Facebook messages [11]. Each minute, social media gains 840 new users [12], YouTube users watch 4,146,600 videos and Instagram users upload 46,740 million posts. Artificial intelligence schemes and Support Vector Machines [13,14,15] focused on finding particular text strings, can be thwarted by the move toward image, and even audio, content on new platforms such as Instagram and Snapchat where, furthermore, content can be set to automatically disappear within a few minutes. There is also the issue of encryption, with sites such as WhatsApp and Telegram becoming a preferred instrument.

Current strategies for tackling online hate etc. tend toward two ends of the size spectrum: The microscopic approach aims at trying to identify some overriding 'bad' individual(s) within the vast sea of online users, which is the quintessential needle-in-the-haystack problem [16]. By contrast, the macroscopic approach is to simply blanket ban entire collectives associated with a particular ideology and hence face allegations of stifling free speech, e.g. Facebook's blanket banning of Ku Klux Klan (KKK) ideology [17]. In the context of complexity science, the equivalents of these two approaches when trying to understand a phase transition [18] for example, are the microscopic viewpoint of looking for the 'bad' particle in a sea of billions (even though there isn't one) or the macroscopic viewpoint that the entire system is to blame (akin to thermodynamics). However, it is known that the correct science lies at the *mesoscopic* scale, in the dynamics of the clusters of correlations that develop near the transition point [18].

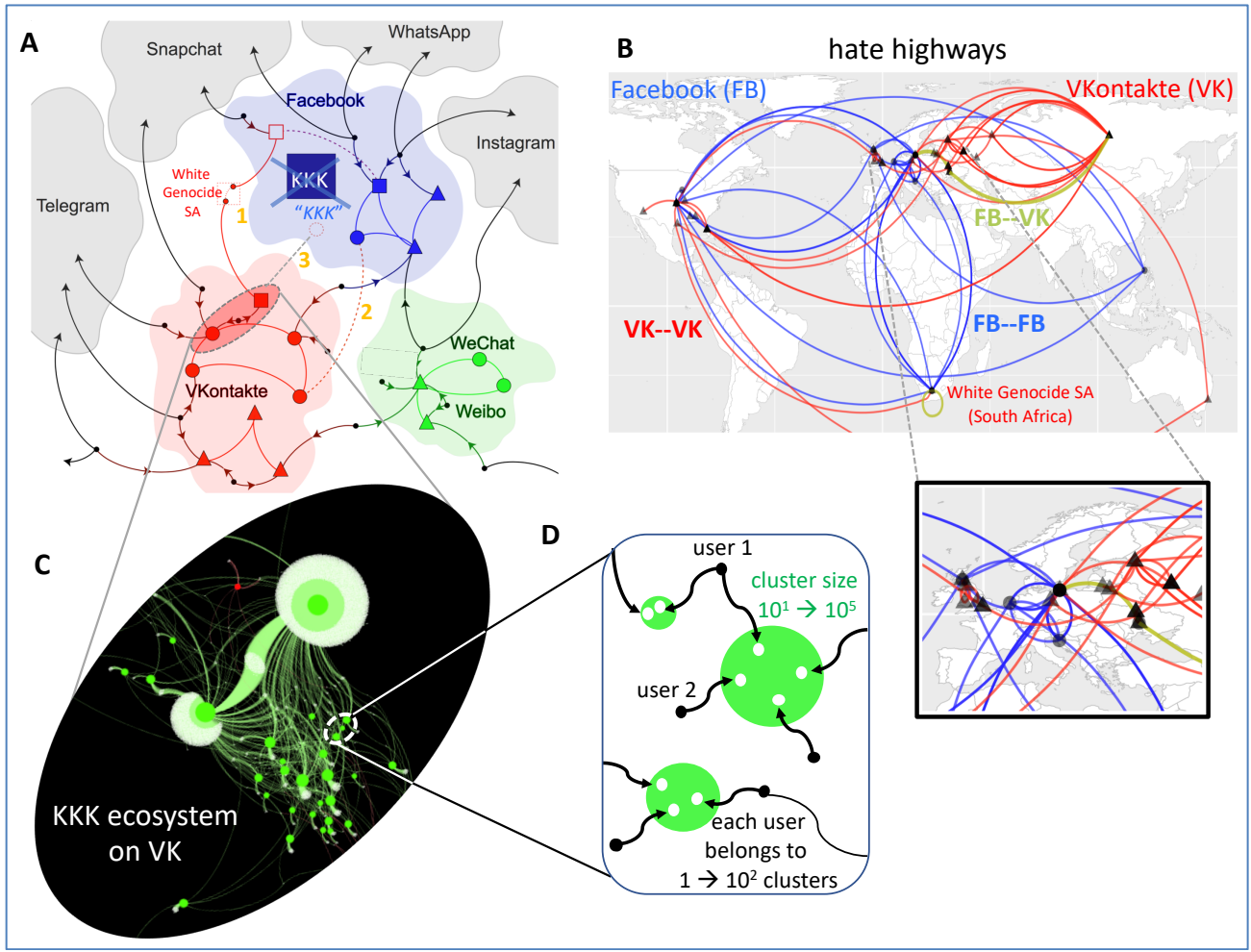

Figure 1. The online hate universe. A: Hate map indicates current online universe of interconnected social media platforms, e.g. Facebook (FB), VKontakte (VK). Though incomplete because of limited legal access to all platforms' data, it captures schematically the key features that we have observed from our analysis of the empirical data. Hate clusters (e.g. FB and VK groups) are shown as larger symbols with links between distinct symbols indicating how different foci are interconnected. Small black circles are users, who may be members of 1,2,3… hate clusters. The data reveals the three types of inter-platform connection that emerge but lie well hidden in the system: (1) hate-cluster mirroring, (2) direct inter-hate-cluster linkage and (3) a hate-cluster implant (see Supplementary Material (SM) for explicit examples). B: Hate highways revealed by placing hate-clusters at the approximate geographic location of their activity, e.g. Stop White Genocide located in South Africa. Links between hate-clusters build highways of hate. When the hate-cluster's focus is an entire country or continent, the geographical center is chosen. Inset shows that particularly dense highways emerge within Europe, between U.K., Holland, Scandinavia and central Europe. C: Magnified diagram of the KKK hate-cluster ecosystem on VK, plotted using the ForceAtlas2 layout algorithm, which shows how a striking 'bonding' (akin to chemical covalent bonding) develops organically between two large hate-clusters as a result of shared users. Hate-cluster radii are determined by the number of members, which ranges from a few to tens of thousands. We include KKK ideology because it is classified as hate by the Anti-Defamation League and the Southern Poverty Law Center, and the KKK clusters in our study include material that seems consistent with hate, however we fully acknowledge that their supporters may not agree. Whether a given group is strictly a hate philosophy, or simply shows material with tendencies toward hate, is not a key feature of our study or findings. D: Expanded schematic showing the synapse-like nature of the individual hate clusters.

Figures 1-3 summarize our findings of how online hate evolves through mesoscopic hate-cluster dynamics. The tractability of our minimal yet generative mathematical model in Fig. 2 comes at the expense of missing detail, but can provide a well-defined and quantifiable starting point for policy discussions and 'what-if' interventions for dealing with hate at the system level, within and across social media platforms, and from the local, to national and international level.

Empirically, our focus on the mesoscale dynamics (i.e. cluster behavior) is consistent with the fact that online group-level hate has already been linked to a number of recent violent real-world attacks. Examples include Dylann Roof [19] as well as the Washington D.C. shooter [20], the Maryland attacker [21], the attackers in Charlottesville and Parkland, the pipe-bomber from Miami, the anti-Semitic attacker in Pittsburgh, and the "incel"-inspired attacker in Tallahassee [22]. Previous work has established links between collective online material and eventual real-world terrorist acts [23-26] and also recruiting for such acts [27-29].

Theoretically, our mesoscopic description is consistent with the fact that clustered correlations provide the key to understanding the dynamics of many-body physical, chemical and biological systems [18] and the likely extension to social systems [30-32]. It also builds on foundational work on networks of networks by Kennett, Stanley, Havlin and co-workers [33-35]. In political science and sociology, it is consistent with recent ideas in collective, and specifically connective, action [36,30]. Operationally, it avoids needing access to any detailed information about individual users. Hence it enables progress to be made on this topic without recourse to user-based methodologies that are now curtailed by privacy laws. We also do not need to, and hence do not, mention the specific names of online clusters (i.e. groups and pages) in Figs. 1-3. Instead we simply indicate two generic ideological labels in Fig. 1.

The accessibility of Twitter data has made it very popular for academic studies of hate and extremism [28] as well as fake news, polarization, market manipulation and product promotion [37-41]. However, its open follower structure does not capture the tendency of humans to form into tight-knit social groups of like-minded individuals, within which group members can develop thoughts and plans freely without restrictions on length of content [42] and without running significant risk of being trolled or of encountering opposing opinions. At the other end of the spectrum are platforms like Telegram, that favor privacy for within-group interactions. But this privacy comes at the cost of making it difficult for the groups to broadcast to and attract potential new members. A middle point is offered by social media platforms like Facebook and its counterparts such as VKontakte, which is based in Russia and has >0.5 billion users worldwide. These platforms allow collective accounts (e.g. "Groups" in FB and VK, "Community Pages" in FB, and "Public Pages" in VK) where individual users can coalesce into groups [43,44]. Each group is a unique, well-defined, tight-knit cluster of users with its own URL, name and member list; its own group administrators who are themselves group members; and its own group focus which can then attract new recruits. Moreover, group members are able to collectively self-police other members and report unwanted behavior (e.g. trolling, adverse opinions) to group administrators who can act promptly to remove the relevant users – and group administrators can vet potential new members. Bots can be detected through the frequent interactions between group members and eliminated immediately by the in-group administrators.

The importance of these online groups in influencing people's real-world actions has been demonstrated recently in a number of settings, e.g. stay-at-home-dads (SAHD), anti-vaccinations, anti-pregnancy, and alternative medications for treating cancer [45-48]. Groups provide a means for developing narratives, support, funding and recruitment, and do so in a way that feels safer and more intimate to their users [49]. Indeed, encouraging group formation and function by adding additional tools for group administrators and members, was recently announced as a primary mission of Facebook and is likewise being copied by other social media companies [50]. However, just as these powers can protect innocuous groups from harm, they can also protect groups focused on hate. This can make joining such a group that promotes hate attractive since it significantly reduces the risks of being trolled or being confronted by opponents. Even on platforms that do not formally have groups, quasi-groups can now be formed: on Telegram these are in the form of groups and super-groups [28,51]. Hence our analysis of cluster dynamics is likely applicable to all other present, and future, social media platforms. Telegram also allows the formation of pods that can then feed into Instagram [52], and also Snapchat [53]. Importantly, the latter two platforms are likely the most relevant for younger generations who tend not to use Facebook [54]. Even though groups can formally be open (i.e. everyone sees members and content), closed (i.e. see members but not all detailed content), or secret (i.e. like closed but cannot be searched), most are open since they aim to attract new support with the hope that they remain hidden by the sea of online activity.

We focus here on such groups on Facebook and VKontakte, as well as "community/public" pages which are effectively a weaker form of group structure. Our snowballing methodology is the same as in our earlier work on pro-ISIS online activity [29,55-57] (SM). Groups and pages both allow humans to cluster, and for simplicity are both referred to here as **clusters**. At no stage do we need to enquire about the real identities or personal information of the users, just as understanding how water boils only requires a description of clusters of correlated molecules [18]. Instead, our analysis is confined to the mesoscale cluster dynamics that are freely visible on these platforms. Our study also serves to shed light on the online world that potentially vulnerable individuals, including minors, will experience and absorb when they happen to hit upon one of these groups and then follow its links.

Figure 1A shows an illustrative portion of the global wiring-diagram of hate clusters for a representative recent day (October 1, 2018). Our results in this paper focus on the largest clusters around topics of neo-Nazi white supremacy, anti-Semitic, anti-Muslim, anti-LBGQT etc., however a full map should eventually be developed to include all forms of hate and all platforms. Starting with a given hate cluster A, we look outward to find any cluster B to which cluster A is explicitly and strongly connected, as opposed to simply having a few common members. We also developed software to perform this process automatically and, upon cross-checking the findings with our manual list, we were able to obtain approximately 90% consistency between the manual and automated versions. We iterated this process until closure of the list (i.e. the search led back to clusters that were already in the list) on a daily basis in real time, yielding a few hundred hate clusters. Highways exist between hate clusters across platforms, countries, languages and particular ideologies (Fig. 1B). Instead of hate residing in 'corners' of the online social media world, we find that it favors a high-dimensional structure that crosses boundaries of physical platform, geographic location, nationality, language and particular ideology (Figs. 1A and 3). For example, neo-Nazi clusters associated with English football, mix white supremacist skinhead imagery while promoting Black music genres. This eclectic and internally contradictory mix of themes makes the global web of hate like an all-encompassing fly-trap that can quickly capture new recruits without them yet having a clear focus for their hate.

The resulting global hate network (Fig. 1A) represents a novel example of a large-scale, *temporal bipartite network-of-networks* whose complex structure evolves dynamically over time as a result of the ongoing interplay between (1) the addition of connections from individual users (black dots) into a given hate cluster (larger colored shapes) when the users join that cluster, or the removal of links when the users leave; (2) the addition of links between individual hate clusters when they link to each other; (3) the addition or removal of connections between platforms. Even though platforms such as Facebook and VKontakte are operationally and financially independent social media networks -- and likewise the clusters within each platform are nominally independent in that they can have different administrators and members, names and locations – we observe three different types of robust inter-platform connections (see Fig. 1A and caption, with specific examples in the SM).

We also observe a remarkable adaptation strategy across platforms, in which a banned ideology on platform A first migrates to platform B – and then when chased from platform B, reappears at a later state on platform A, where it is still banned, but now in a 'reincarnated' form that is harder to catch. Figure 1A shows the specific example of the Ku Klux Klan ideology, which became banned from Facebook and instead evolved an ecology of clusters (nearly 60) on VKontakte. These included some Ukranians as cluster members: When Russia invaded Ukraine and VKontakte became banned [58] this ecosystem migrated back to Facebook *but* profited from using titles in Cyrillic which made them likely harder to catch by platform A's machine-learning detection algorithms.

Figure 2 presents a mathematical theory for exploring the combined impact on the global online hate universe of different platforms separately trying to remove hate content. It considers hate clusters and their members in one platform (e.g. platform 1) incurring a cost $R$ to access another platform (e.g. platform 2). This cost $R$ mimics their risk of exposure and hence potential sanctions in the form of suspension of users' accounts or possible legal action. For concreteness, we consider the common situation that we find in the empirical data, of a subset (1b) of $c$ hate clusters on a given social media platform forming a loop. While each hate cluster may have links to other clusters outside the subset, we find that these invariably lead down a rabbit-hole, and hence do not form additional simple loops. We choose the schematic of Fig. 2A which is a simplification of Fig. 1A, however we stress that our theory can be applied to any loop-like arrangement of clusters, any form of cost function $R$, and any arrangement of social media platforms. Building on the mathematics of Ref. 59 and in the SM, and assuming a simple probability $q$ of a given hate cluster making a link to another platform, we can derive a formula for the path-length $l$ of the highway (i.e. shortest path) between any two hate clusters in subset 1b on platform 1. The resulting formula for the average value of the shortest path (i.e. length of hate highway) between any two hate clusters in Fig. 2A, and hence also in Fig. 1A, is then given by

$$\bar{\ell}(q,c,R) = \frac{R(R-1)}{2(c-1)} + \frac{(1-q)^{c-R}\bigl(3+(c-2-R)q\bigr)}{q^2(c-1)} + \frac{q\bigl(2-2R+2c-(R-1)(R-c)q\bigr)-3}{q^2(c-1)} \quad (1)$$

This same formula applies for any number $c$ of hate clusters and any cost $R$. Apart from an unimportant scaling factor, it also applies even if the strength of support between clusters A and B, B and C etc. is highly asymmetric. Though $R$ is in principle general, we consider here the cost $R$ increasing linearly with the number of links between the two platforms, which is consistent with the idea that more links between platforms 1 and 2 will make platform 1 more noticeable to platform 2's managers and law enforcement. Figure 2B shows that the average shortest path in Eq. 1 as a function of the number of links between platform 1 and 2, then has a minimum -- and gives approximate expressions for this minimum as obtained from Eq. 1.

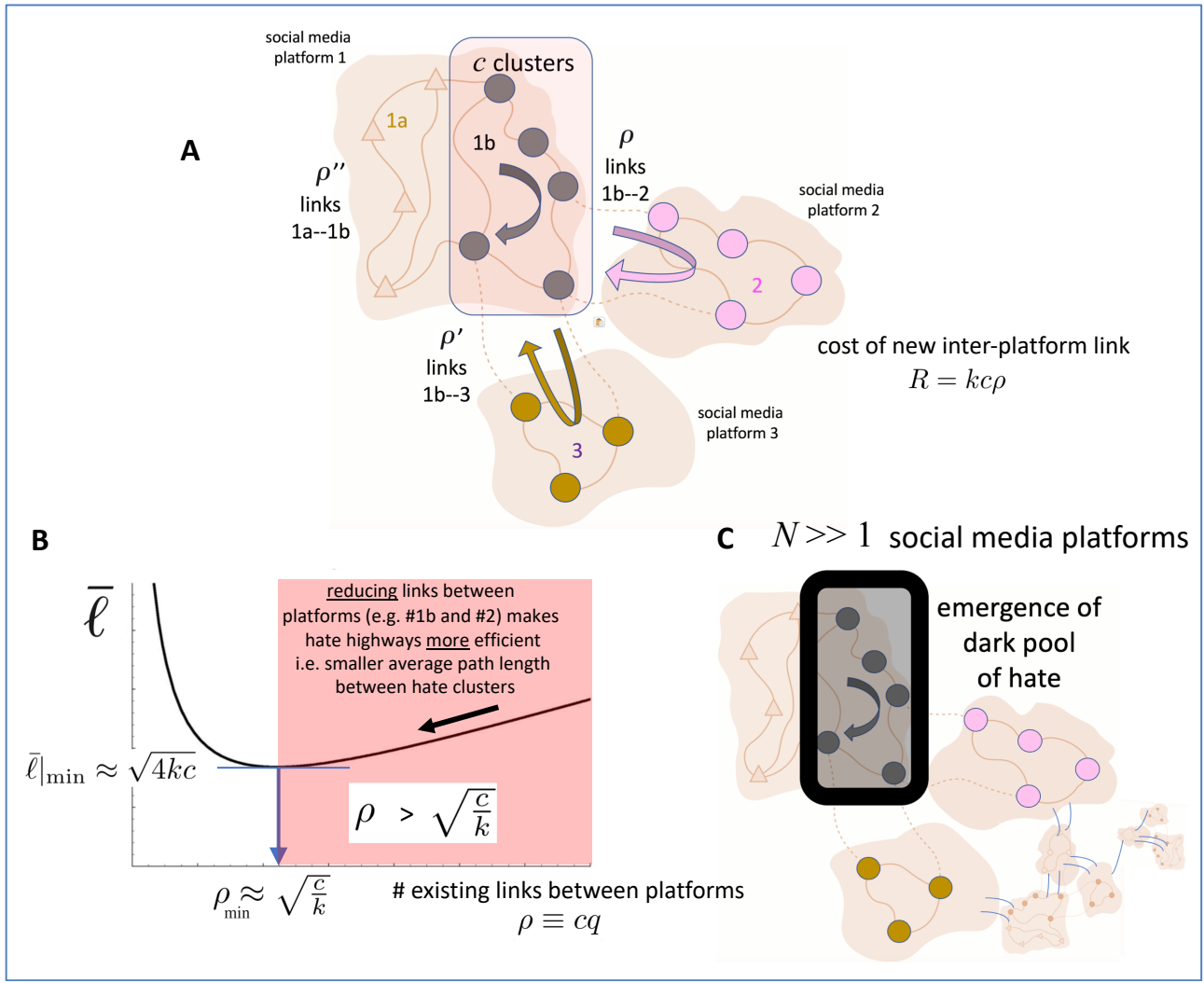

**Figure 2: Mathematical theory of multi-platform hate highways. A:** Establishing links (as observed empirically, see Fig. 1A) from a platform like VK (1b) to a better-policed platform like FB (2) runs the risk (cost $R$) of being noticed by FB moderators, and hence sanctions and possibly even legal action. Since more links creates more risk, we choose $R$ as proportional to the number of links with $k$ being a constant of proportionality. **B:** Theory shows that counterintuitively, the diligent action of moderators on platform 2 (e.g. FB) in reducing the number of inter-platform links $\rho$, unwittingly reduces the shortest path between clusters in platform 1b, hence making the hate highways between clusters in 1b stronger and more efficient (as measured by the shortest path length, which in turn is relevant for sharing hate material and messaging). Though we choose panel A here to resemble Fig. 1A, the same theory (and hence Eq. 1, derived in the SM) applies to any of the loop-like arrangements of clusters that are commonly observed in the data (see SM for empirical example). **C:** As the universe of social media expands in the future to many interconnected platforms, our theory predicts that the combined effect of having diligent yet independent moderators on each platform will create spontaneous dark pools within which there are superhighways of hate. A formidable adversary can then form within these dark pools, and can subsequently launch a strong attack on other platforms (e.g. FB). The theory also predicts that by coordinating moderators' actions across platforms, the corresponding costs $R$ (See Eq. 2) can be chosen in a coordinated way to significantly reduce the likelihood of this undesirable outcome.

The counterintuitive policy implication of our mathematical theory, and hence Fig. 2B, is that a seemingly sensible approach by a moderator on platform 2 to decrease the number of links between platforms 1 and 2, can actually lead to an *increase* in the hate highway efficiency within platform 1 (i.e. decreased distance between any two hate clusters in platform 1) which in turn generates a stronger community of hate on platform 1 – and hence ultimately generates a more potent threat facing platform 2. Employing renormalization, Eq. (1) can be generalized to describe loops-of-loops and hence applied across different levels of structure within and between platforms. The length of the highway between any two hate clusters given the full hate universe in Fig. 1A with $i$=1,2,..$N$ platforms, then becomes:

$$\bar{l}_{i+1}(q_{i+1}, c_{i+1}, R) = \frac{(1-q_{i+1})^{c_{i+1}-\bar{l}_i(q_i,c_i,R)}\left(3+(c_{i+1}-2-\bar{l}_i(q_i,c_i,R))q_{i+1}\right)}{q_{i+1}^2(c_{i+1}-1)}$$
$$+\frac{2q_{i+1}\left(1-\bar{l}_i(q_i,c_i,R)+c_{i+1}\right)}{q_{i+1}^2(c_{i+1}-1)} - \frac{q_{i+1}\left((\bar{l}_i(q_i,c_i,R)-1)(\bar{l}_i(q_i,c_i,R)-c_{i+1})q_{i+1}\right)-3}{q_{i+1}^2(c_{i+1}-1)}$$
$$+\frac{\bar{l}_i(q_i,c_i,R)(\bar{l}_i(q_i,c_i,R)-1)}{2(c_{i+1}-1)} \quad (2)$$

It can be shown mathematically (see SM) that a gradual, monotonic change in the cost parameter $R$ can now lead to an *abrupt* change in the functional behavior of the online hate universe, with hate highways abruptly re-routing to bypass unwilling platforms, and potentially strengthening their existing inter-cluster links in the process. This will unfortunately create sudden, out-of-the-blue surprises for hate moderators at the global level. In addition, in an online universe with many social media platforms (Fig. 1A), Eq. 2 identifies a hidden resilience in the global axis of hate as illustrated in Fig. 2C, whereby hate clusters in a given platform 1 can spontaneously and rather abruptly emerge as a 'dark pool' (1b). This dark pool retains links to other hate clusters in platform 1 and to platform 2 but these are not used, hence monitoring of platform 1 from platform 2 gives the false impression that platform 1 is innocuous – and similarly for any monitoring of subnetwork 1b from within the rest of platform 1. This hidden resilience offers an explanation as to why current control methods of banning an entire sector of clusters on a given platform, are likely to be ineffective. Fortunately, Eq. 2 also predicts that by coordinating moderators' actions across platforms, the corresponding costs $R$ can be chosen in a coordinated way to avoid this undesirable outcome.

Figure 3 reveals a remarkable bottom-up rewiring and repair capability that lies hidden in the online hate ecosystem at finer resolutions, as revealed by its response following the 2018 Parkland school shooting. While we know of no evidence that KKK was involved in this, it is known that accusations and conspiracies did circulate in the days following the attack, and Fig. 3A shows specifically how the KKK clusters on VKontakte responded. In particular, links akin to chemical 'bonds' form between previously unconnected KKK clusters. We speculate that this is an adaptive evolutionary response by which the decentralized KKK ideological organism manages to come together to protect itself by bringing together previously unconnected supporters in a collective way. The remarkable aspect is that this appears to happen in a bottom-up, organic way, increasing the cluster-projected network clustering, degree and hence togetherness (Fig. 3B). Figure 3C shows the network on a larger scale around February 17, with the bonding density of common users clearly visible (white cloud between the green clusters). Interestingly, similar bonding is also seen in the response of anti-Western jihadi hate groups during 2015 [60] when the leader of ISIS was reportedly injured in an air strike (Fig. 3D), which leads us to speculate that the 'covalent bonding' is a general feature. Also noteworthy is the fact that this dynamical KKK bonding conserves the distribution of KKK group sizes over time, with each day producing a similar distribution to Fig. 3E (left panel). The approximate power-law has a high goodness-of-fit $p$=0.92 and the power-law exponent 1.7 is consistent with our data collection process of extracting cluster sizes from across the different KKK cluster ecologies on VKontakte (see SM). Specifically, the SM shows mathematically and using simulation that aggregating data from a set of distinct power-law distributions having exponents near 2.5, as expected based on previous studies of jihadi online groups [29], yields an approximate power-law with exponents in the range 1.7-1.9, as we indeed observe in the empirical data (Figs. 3E,F).

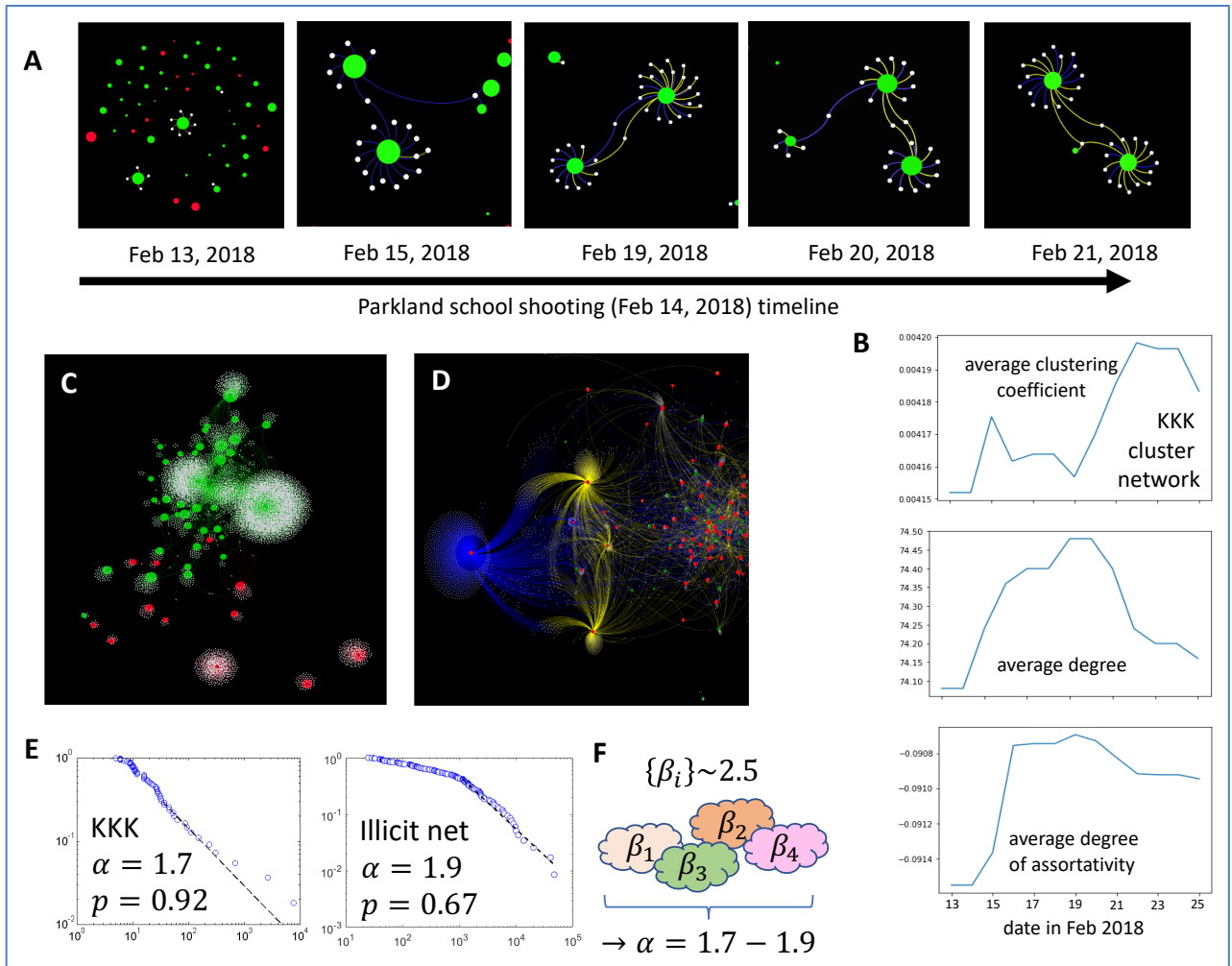

**Figure 3. Online hate cluster ecosystem rewiring and self-repair. A:** KKK ecosystem on VKontakte prior to and after the February 14, 2018 Parkland school shooting. During the subsequent week, the organic rewiring served to create a 'bond' between previously distinct KKK clusters. Such a need to pull together, coincides with a period of uncertainty for the KKK that was generated by conspiracy theories about culpability. For clarity, only users (white circles) that change their status in the next timestep are shown. Larger red nodes are clusters that are closed (i.e. closed VK groups), green nodes are open (i.e. open VK groups). Yellow links mean the user will leave the cluster between day $t$ and day $t+1$, meaning that link is going to disappear. The blue links mean the user joins the group on day $t$. **B:** This pulling-together is confirmed by network measures for the KKK cluster network. **C:** Full KKK network snapshot 1 week after Parkland shooting, confirms the establishment of a 'bond' between the largest KKK clusters that did not exist before. **D:** A remarkably similar 'bonding' effect arises in the jihadi hate cluster ecology associated with pro-ISIS support, around March 18, 2015 a few days after a coalition strike appears to have wounded ISIS leader Abu Bakr al-Baghdadi. Rumors immediately circulated among the clusters that top ISIS leaders were meeting to discuss who would replace him if he died, suggesting that his injuries were serious. However, none of this become public knowledge in the media, i.e. the rewiring and self-repair leading to conversion of two clusters into one (i.e. two disappeared, shown yellow, and one appeared, shown as blue) indeed appears to be bottom-up, organic. While C mimics electronic covalent bonding, D is a more extreme version of bonding akin to nuclear fusion. E: Both the KKK ecology and the ecology of illicit financial activities, show support for a power-law distribution with exponents between 1.7 – 1.9 and high goodness-of-fit *p*-values [61]. F: This range of power-law exponents in E is consistent with aggregating data from different thematic subsystems, each of which has power-law distribution with exponent broadly distributed around 2.5 as found earlier in Ref. 29. See SM for full analysis.

In addition to highlighting the need to coordinate moderation across platforms in order to avoid creating hate dark pools (Fig. 2C), our findings suggest that platforms can benefit by avoiding white-label banning (e.g. KKK in Fig. 1A) and instead look to influence the 'bond' dynamics that arise following particular real-world events (Figs. 3A-D). There are of course many limitations of our study, most notably that we do not yet have a complete online map (Fig. 1A) due to the current lack of access to platform data that faces both researchers and government agencies [62]. Despite their incompleteness, we believe our findings can nonetheless also shed light on other illicit networks which operate under similar pressures, i.e. networks that are simultaneously open enough to find new recruits/victims and yet sufficiently hidden that they can avoid capture. Using recent data obtained from online clusters pursuing financial fraud [63] we indeed find preliminary quantitative similarities: e.g. the distribution of cluster sizes is again an approximate power-law with exponent in the range 1.7-1.9 (Figs. 3E and F).